\documentclass[twocolumn,showpacs,pre]{revtex4}

\usepackage{graphicx}
\usepackage{mathrsfs}

\begin{document}

\def\bq{{\mathbf q}}
\def\br{{\mathbf r}}
\def\bu{{\mathbf u}}
\def\bv{{\mathbf v}}
\def\bw{{\mathbf w}}
\def\bx{{\mathbf x}}
\def\by{{\mathbf y}}
\def\bz{{\mathbf z}}
\def\bA{{\mathbf A}}
\def\bD{{\mathbf D}}
\def\bI{{\mathbf I}}
\def\bJ{{\mathbf J}}
\def\bQ{{\mathbf Q}}
\def\bS{{\mathbf S}}
\def\bV{{\mathbf V}}
\def\bW{{\mathbf W}}
\def\mD{{\mathcal{D}}}

\title{General representation of collective neural dynamics \\
  with columnar modularity}
\author{Myoung Won Cho}
\email{mwcho@postech.edu}
\author{Seunghwan Kim}
\email{swan@postech.edu}
\affiliation{
  Asia Pacific Center for Theoretical Physics $\&$ NCSL,
  Department of Physics, Pohang University of Science and Technology,
  Pohang, Gyeongbuk, 790-784, Korea}
\date{\today}
\begin{abstract}
We exhibit a mathematical framework to represent the neural process at the
cortical level.
The description of neural dynamics with columnar and functional modularity,
named the fibre bundle map (FBM) representation method, is based on both
neuroscience and informatics, whereas it leads to the conventional formulas in
statistical physics.
The possibility of analogy between the phenomena in brain and physical systems
has been proposed~\cite{Cho2004A,Cho2004B}.
In spite of the complex circuitry and the nonlinear dynamics in neural systems,
the neural behavior at high levels may be described by simple and general
rules, which related to the noble theory in statistical physics.
The FBM method gives profit in building or analyzing the neural models by
representing essential ingredients of neural interactions by general formulas.
%We insist that the typical characters of self-organizing map is determined not
%by the detailed and complex interaction rules but by the topology of lattice
%and feature space.
%The collective neural phenomena can be understood and predicted through some
%parameters in a general energy form with symmetry transform invariance.
%Not only the similarity in formulas, the cortical dynamics can share the
%statistical properties with other physical systems, which validated in primary
%visual maps~\cite{Cho2004A}.
We apply our method to the proposed models of visual map formation and show how
they can share statistical properties with vortex dynamics in magnetism in
spite of various development mechanisms.
\end{abstract}

\maketitle

\section{Introduction}
Though the detailed dynamics of a single neuron are revealed,
there still remains a challenge at the network level to explain how brains
perform higher cognitive functions.
Studies of physical models are focused on achieving a biological realism of the
neural computation models.
%such as the networks of coupled oscillators~\cite{Nishikawa2004,Aonishi1999},
However, the success of the basic neural network models, based on the
connectional framework between simple cells, in the application of small
adaptive systems, they are inherently problematic in the apprehension of
collective neural phenomena and higher cognitive behavior in the real brain.
And also there are attempts to see through the neural processing at higher
levels, the functional modularity of neurons or the symbolic processing
architecture.
Before the physiological evidence of repetitive cortical blocks, there were
proposals of the modularity within neighbor neurons, the called {\em cell
assemblies} (CAs), considering the high dimensional attribute and faculty of
neurons~\cite{Hebb1949}.
It is a tendency of neurons to aggregate together with similar functional
specializations and make organizations hierarchically.
Though different classifications and names for neural clusters, we adopt the
suggestion that {\em neuron} - {\em minicolumn} - ({\em hypercolumn}) -
{\em macrocolumn} - {\em cortex area} - {\em hemisphere}, where minicolumn is
a candidate for ``the repeating pattern of circuitry'' or ``the iterated
modular unit''~\cite{Calvin1998}.

In this paper, we exhibit a mathematical framework, noted briefly and named the
fibre bundle map (FBM) methods in ref.~\cite{Cho2004A}, and show how to
represent the neural process generally at the cortical level.
Briefly speaking, the FBM representation is a mapping of the feature components
(often the synaptic weights) to topological spaces, the called {\em bundles}.
The pattern informations are reordered in locally transformed coordinates and a
few of major components are extracted.
Obviously there exist another mathematical framework to represent the neural
process in reduced space.
Kohonen set up the mathematical preliminaries, the called feature maps or
{\em feature-based} representation, in vector space and led the successive
models in artificial and physiologic neural networks~\cite{Kohonen1984}.
Symbolic processing architectures also suggest the description of neural
computations at the cognitive and rational bands.
The feature vector space or the symbolic sets can belong to a kind of FBM
representation.
But the FBM method has an interest in the manifold structure of frequent
inputs in feature space and its corresponding symmetry group.
Indeed, the properties of neural progress are governed not by the detailed
neural interaction rules but by the algebraic structure of dominant feature
components.
The FBM method represents the neural process by the general formulas in
statistical physics and helps to comprehend collective neural phenomena
intuitively through the knowledge in statistical mechanics and differential
geometry.

As a class of abstract representations, the formulas in the FBM models have
some different character with those in the feature-based (often the called
``low-dimensional'' feature vector) models.
In the feature-based representation, the change in the feature vector at
position $\br$, $\Phi(\br)$ is described as the difference in the stimuli
vector $\Phi(\br')$, such as $\Delta\Phi(\br)\propto(\Phi(\br')-\Phi(\br))$
with the energy of the form $|\Phi(\br)-\Phi(\br')|^2$ (or its higher powers).
%Whereas, the energy functions in the FBM representation consist of the inner
%products such as $\psi(\br)\psi'(\br)$.
Whereas, in the FBM representation, the interactions between neurons are
notated by inner products rather than their distance, and it is generally
assumed that the energy of neural process can be expanded in a power series,
i.e.,
%and classified according to the number of coupling.
%It is assumed that the energy of neural process can be expanded in a power
\begin{eqnarray} \label{eq:expansion}
  E[\psi]&=&E^{(0)}-\sum_iB(\br_i)\psi(\br_i) \nonumber \\
  &-&\frac{1}{2!}\sum_{i,j}D(\br_i,\br_j)\psi(\br_i)\psi(\br_j) \\
  &-&\frac{1}{3!}\sum_{i,j,k}F(\br_i,\br_j,\br_k)
    \psi(\br_i)\psi(\br_j)\psi(\br_k) + \cdots. \nonumber
  %&-&\frac{1}{4!}\sum_{\bx,\by,\bz,\bw}G(\bx,\by,\bz,\bw)
  %  \psi(\bz)\psi(\by)\psi(\bz)\psi(\bw)+\cdots, \nonumber
\end{eqnarray}
where the field variables $\psi(\br)$ denote the feature state of neurons at
cortical location $\br$.
We will show how this formula is derived and the interaction functions are
determined in the simple (or the called ``high-dimensional'' feature vector
representation) and the complex cell models.
%This formula can be derived concretely from the fundamental neural process as well.
%Moreover, in the simple cell (or the ``high-dimensional'' representation)
%models, the objective function for Hebbian modification obey this general form
%as a function the synaptic weights $\bW$ rather than the field variables $\psi$.
In a continuum limit, the energy can be approximated to
\begin{eqnarray} \label{eq:continuum}
  E[\psi]=\int d\br\left\{\frac{v}{2}|(\nabla-i\bA)\psi|^2
    +\frac{m^2}{2}|\psi|^2+\frac{g}{4!}|\psi|^4\right\},
\end{eqnarray}
where the odd power terms are expected to be vanished generally.
This is just the Ginzburg-Landau energy with gauge invariance and explains the
statistical properties of the emergent cortical maps in experiments and
simulations.
The energy in a continuum approximation often can be derived using only minimal
mathematical constraint such as the symmetry.
%requirement of invariance under the symmetry transformations without the detailed cortical modification rules.
The energy form in Eq.(\ref{eq:expansion}) and Eq.(\ref{eq:continuum}) proposes
the possibility of the analogy between the physical and the neural systems, and
the characteristics of developed visual maps are systematically apprehended
through the statistical properties of vortices in
magnetism~\cite{Cho2004A,Cho2004B}.
%Phase transitions can be predicted when the changes in parameters, whereas the
%parameters are obtained from the detailed interaction mechanisms.

%The general energy form in cortical dynamics can be build via two different
%ways.
%One, the energy function and the pattern properties in cortical map formations
%can be inferred only using the topologic properties.
%the symmetry between the feature states (or called {\em gauge symmetry} in quantum mechanics).
%Considering the transform invariant properties, it is generally assumed that
%the energy of map formations takes the form at a continuum limit
%Another way is to build models through the detailed description of individual
%neural interactions.

We apply the FBM representation method to the development models in visual
cortex.
The cortical map formation in orientation and ocular dominance columns is one
of the most studied problems in brain.
A considerable amount of different models is proposed, and some of which are
compared with the experimental findings and in
competition~\cite{Erwin1995,Swindale1996}.
%The theoretic analysis of pattern formation are reported within a few of models.
Miller {\em et al.} formulated {\em correlation-based} models describing how
ocular dominance and orientation columns develop in simple cell
models~\cite{Miller1989,Miller1992,Miller1994}.
Obermayer {\em et al.} presented a statistical-mechanical analysis of pattern
formation and compared predictions quantitatively with experimental data using
the Kohonen's {\em self-organizing feature map} (SOFM) approaches.
Wolf {\em et al.} obtained again the conditions for the emergence of a columnar
pattern in the SOFM algorithm~\cite{Wolf2000}.
The studies of the {\em elastic-net} model also show the bifurcation and
emergence of a columnar pattern~\cite{Durbin1990,Hoffsummer1995,Goodhill2000}.
Scherf {\em et al.} investigated the pattern formation in ocular dominance
columns with more detailed model, which covers the results of the SOFM
algorithm and the elastic-net model~\cite{Scherf1999}.
Wolf and Geisel predicted the influence of the interactions between ocular
dominance and orientation columns on the pinwheel stability without model
dependency and demonstrated it in the simulations of the elastic-net
model~\cite{Wolf1998}.
The lateral (or neighbor) interaction models are also successful scheme based
on physiology~\cite{Swindale1980,Swindale1982,Cowan1991,Cho2004A}.

In the proposed visual map formation models, the Hamiltonian models with spin
variables belong to the class of FBM representation
models~\cite{Cho2004A,Cowan1991,Tanaka1989}.
Other development models written in the high- or low-dimensional feature vector
representation can be revised again in the FBM representation.
The formulas in FBM models represent essential ingredients of neural
interactions without paying much attention to particular neural control
mechanism.
Moreover, the modification of the iterative procedure of a model into the
formula in Eq.(\ref{eq:expansion}) or Eq.(\ref{eq:continuum}) becomes the
statistical analysis of the model itself.
The quadratic interaction function $D(\br_i,\br_j)$ is consequence in the
visual map formation as other physical systems.
The interaction functions in neural process mean more than the intracortical
connections or recurrents in the called lateral activity control.
In the competitive Hebbian models, such as the elastic-net model and the SOFM
algorithm, the interaction functions comprise the feedforward competition or
normalization process.
However, in the FBM representation the functional matrix $D(\br_i,\br_j)$ of
the visual map formation models have common shape, the called Mexican hat type,
that is, positive in short-range and negative in long-range, in spite of
different development mechanisms.
The bifurcation to a inhomogeneous state and the emergence of a columnar
pattern is possible when there are strong negative interactions in
$D(\br_i,\br_j)$.
The development of a columnar pattern is also concerned with non-vanishing
vector $\bA$, the called {\em vector potential} in physics, in
Eq.(\ref{eq:continuum}).
The FBM representation method will show how the development models with
different mechanisms lead to the successful formation of visual maps and
share the statistical properties of vortices in the spin Hamiltonian models.
%Recently, we predicted the bifurcation of inhomogeneous solutions also in
%lateral interaction models, and derive the typical properties in observed
%patterns, such as the orthogonality and the correlation function~\cite{Cho2004A}.

\section{Representation of neural state with columnar modularity}
The structures and connections in cerebral cortex are more complex and modular
than those in artificial neural networks.
Neurons tend to be vertically arrayed in the cortex, forming cylinders known as
cortical columns. 
Traditionally, six vertical layers have been distinguished and classified into
three different functional types.
The layer IV neurons ({\em IN} box), first get the long-range input currents,
and send them up vertically to layer II and III ({\em INTERNAL} box) that are
the called true association cortex.
Output signals are sent down to the layer V and VI ({\em OUT} box), and sent
further to the thalamus or other deep and distant neural structures.
Lateral connections also occur in the superficial (layer II and III) pyramidal
neurons.
In columnar (or horizontal) clustering, there are minicolumns, which are
consisted of about 100 neurons and 30 $um$ in diameter in monkeys, and
macrocolumns, which are 0.4$\sim$1.0 $mm$ and contain at most a few hundred
minicolumns.
On the wider discrimination, there are 52 cortex areas in each human
hemisphere; a Brodmann area averages 21 $cm^2$ and 250 million neurons grouped
into several million minicolumns~\cite{Calvin1998}.

%\begin{figure}[t] 
%\includegraphics[width=8cm]{3d-colmn}
%\caption{ \label{fig:3d-colmn}
%  The 3-D structure of cortical columns.
%  (Reprinted by permission from William H. Calvin, 2001,
%  {\em The Cerebral Code}, The MIT Press, Copyright \copyright 1996 by William
%   H. Calvin)
%}
%\end{figure}

The columnar modules can be regarded as a kind of multi-layered neural networks
and would have complex functional attributes.
Most neurons in brain have the attribute of {\em selective response} to a
received activity, and the preferred signals become an useful representation
of the functional attributes in a small neural region.
A traditional representation of neural state is the vector notation $\bv$,
where its components correspond to the activity of each neuron in receptor
layer.
If a columnar module (or complex cell) at position $\br$ respond selectively
to a particular input vector $\bv$ and make an output vector $\by$, its
functional attribute can be represented compactly as,
\begin{eqnarray} \label{eq:associator}
  w(\br)=\by\circ F\circ\bv^\top,
\end{eqnarray}
where $F$ is the nonlinear response or activation function of complex cell.
% posterior probability function.
If the activation function is linear or ignored, this leads to a simple pattern
associator, the called {\em linear associator}.
The experiments of the response properties to external stimuli through
electrode penetration can be understood as the measurement of the product
between the associator $w(\br)$ and the input signal $\bv'$ :
\begin{eqnarray} \label{eq:inner_product}
  |w(\br)\circ\bv'|=|\by|\ F(\bv^\top\bv'),
\end{eqnarray}
where the activity of the output $|\by|$ corresponds to the measurement of the
number of action potentials or the frequency of spikes.
In the physiological experiments with the complex cells in primary visual
cortex~\cite{Hubel1962} or the object perceptions in inferotemporal (IT)
cortex~\cite{Tsunoda2001}, the response property of columnar modules used to be
the combination of different patterns and then the functional form in
Eq.(\ref{eq:associator}) would be expanded into the summation of associators.
When the output $\by$ is common with the most favorite input $\bv$ such as
Hopfield networks~\cite{Hopfield1982} or the most favorite input is only
concerned, a vector notation can play the role of representation of functional
attributes in columnar modules.

%\begin{figure}[t] 
%\begin{minipage}[b]{4cm}
%  \includegraphics[width=4cm]{intrinsic} (a) intrinsic type
%\end{minipage}
%\ \ 
%\begin{minipage}[b]{4cm}
%  \includegraphics[width=3.3cm]{extrinsic} \\ \ \\ \ \\ (b) extrinsic type
%\end{minipage}
%\caption{ \label{fig:coding_type}
%  For the response properties of neurons, two different encoding types are
%  possible whether the synaptic connections are (a) between close neurons
%  within columnar module or (b) with far aparted neurons cross cortex areas.
%}
%\end{figure}

\begin{figure}[t] 
\includegraphics[width=8cm]{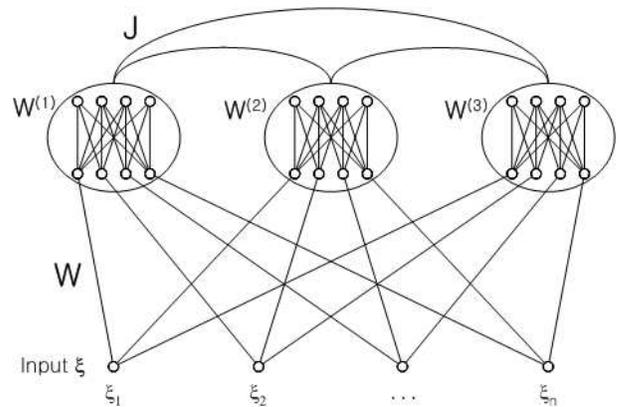}
\caption{ \label{fig:network}
  A neural network model with columnar modules with function $w^{(i)}$.
  Input signal to a columnar module $\bv_i$ is driven by feedforward synapses
  with weights $\bW$, that is, $\bv_i=\bW_i\bu$, and its output $\by_i$ is
  interconnected to neighbors by intracortical connections $\bf J$.
  %Information (or the functional attributes of neurons) are encoded in the
  %connectivity within columnar modules ${\bf w}^{(I)}$ (intrinsic type) or in
  %the feedforward synapses $\bf W$ (extrinsic type).
}
\end{figure}

Fig.\ref{fig:network} depicts a neural network with columnar modules.
A matrix $\bW$ denotes the feedforward synaptic weights cross cortex areas, 
such as the connections between LGN and primary visual cortex, and the input
vector to a columnar module is given by $\bv_i=\bW_i\bu$ (or
$v_i=\sum_\alpha W_{i\alpha}u_\alpha$).
In a complex cell model, it is expected that the synaptic connections within a
columnar module $w(\br)$ achieve the functional attributes of neuron, whereas
in a simple cell model, the connections with the external cells $\bW$ are
considered to vest the functional attributes.
For example, the ocular dominance in primary visual cortex is determined
whether a neuron in V1 is more connected to the left or right eye (or LGN)
cells.
We call this the {\em extrinsic} information coding type, which is realized by
the connectivity of far neurons cross cortex areas, whereas the {\em intrinsic}
type is realized by the synaptic plasticity between close neurons within a
columnar module.
The neural attribute of two coding types are represented by common formula in
FBM models, but there exist some different ground when building actual models.
The feedforward competition behavior should be related to the intrinsic coding
type.
Moreover, the extrinsic encoding type causes a problem in modeling huge
networks because too massive connections are required when the meaning of
activity is characterized only from where the current come.
We expect that the intrinsic type, encoding information in spatial or temporal
correlations within a signal band, is essential in huge networks and would be
a prominent strategy in the real brain.

\section{Fibre bundle map representation}
%In simple cell models, the cortical models can get extremely complex 
% the high-dimensional components the amount of receptor cells,
%To deal with this, a class of more abstract models has been developed.
%In the ``low-dimensional'' feature vector representation each component stands
%for a selected response property.
%For example, the features of orientation columns are denoted by Cartesian
%components
%$\Phi(\br)=\left(q(\br)\sin(2\phi(\br)), q(\br)\sin(2\phi(\br))\right)$ for
%preferred orientation $\phi(\br)$ and degree of preference for that orientation
%$q(\br)$ at each cortical location $\br$~\cite{Swindale1982}.
%In the FBM representation, however, they sometimes takes similar forms with
%the low-dimensional feature vector representation, the feature components are
%approximated with different standpoint.
% given pattern vector, we can extract the feature components,
%that are the center of the pattern $(x,y)$ and the maximal variance vector $(v_x,v_y)$.
%With the ocular dominance $z$, the feature vector with 5 components,
%\begin{eqnarray} \label{eq:visual_feature_vector}
%  \Phi=(x,y,v_x,v_y,z)
%\end{eqnarray}
%is a usual representation of the orientation and the ocular dominance columns in visual cortex.
%so to say, a reduced dimensional
%The components in the FBM representation are composed of 
%representation with the most prominent components on other basis.

\begin{figure}[t] 
\includegraphics[width=8cm]{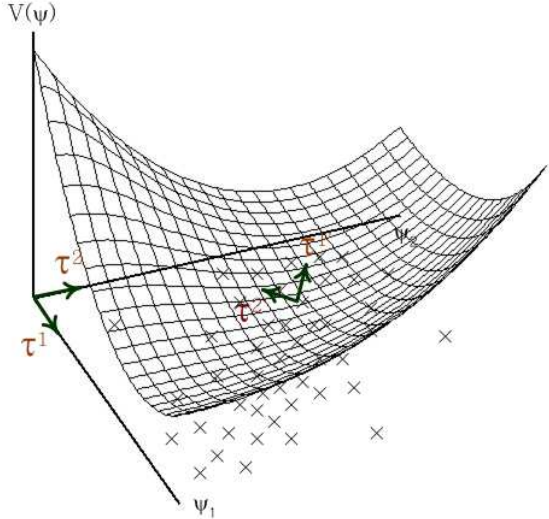}
\caption{ \label{fig:V}
  Probabilistic external stimuli and a potential function with external source.
  The transformed basis $\tau^{1'}$ and $\tau^{2'}$ are the principal
  directions of external stimuli density at a point.
}
\end{figure}

The FBM representation method bases on a mathematical framework - the called
{\em fibre bundle} in manifold theory~\cite{Martin1991,Nash1983}.
For a trivial fibre bundle, a total (or bundle) space $E$, which will depicts
the neural attributes at a cortical area, is composed of a base space $B$ and a
fibre $F$, tat is, $E=B\times F$.
In our interests, cortical locations are the elements in base space, where
feature (often pattern, code or model) space becomes a fibre.
A structure (or symmetry) group $G$ is a homeomorphism of fibre $F$, and the
same with the fibre $F$ in a {\em principal fibre bundle}.
The principal fibre bundles admit {\em connexions} (or vector potentials in
physics), and it is for this reason that they are of basic importance in gauge
theories in physics.
The features of cortical cells or small cortical regions at each cortical
location $\br$ are represented by a set of field variables $\psi_\alpha(\br)$
and
\begin{eqnarray} \label{eq:representation}
  \psi(\br)=|\psi(\br)|\exp(-i\phi_a(\br)\tau^a)=\psi_a(\br)\tau^a,
\end{eqnarray}
where $\phi_a(\br)$ is an arbitrary internal (feature) phase and $\tau^a$ is
the basis of a continuous (or Lie) group G.
The bases can be taken as the amount of receptor cells, but are usually reduced
according to the statistical structure of inputs.
The frequent inputs usually occupy small regions in the total feature space and
the major variance of feature components occurs within a embedded submanifold
with high stimuli density (Fig.~\ref{fig:V})
%, the bases are transformed according to the principal directions of external stimuli density at a point.
The reduction of feature space is related to the extraction of features from
inputs in learning rules as well.
Symmetry breaking between transformed feature components is expected in the
neural progress of experience and learning, and cortical dynamics can be
described with a few of field components in a reduced feature space.

\begin{figure}[t]
\includegraphics[width=6cm]{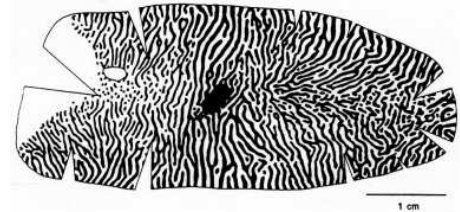}
\caption{\label{fig:macaque}
  The complete pattern of ocular dominance stripes in the striate cortex of a
  macaque monkey.
  There is a strong tendency for the stripes to meet the margin of striate
  cortex at steep or right angles.
  (Reprinted by permission from S.LeVay, Copyright \copyright 1985 by the
  Society for Neuroscience~\cite{LeVay1985}.)
}
\end{figure}

%The interactions in cortical circuitry and the synaptic plasticity are more
The differential geometric concepts in the FBM representation furnish an
intuitive explanation for emergent cortical maps.
The self-organization of feature maps achieved by locally gathering similar
interests means there is smooth variance of features with neighbor neurons at
each location.
In other words, the properties of ``organized'' and ``optimized'' feature maps
is related with those of ``continuous'' and ``flat'' variables in manifold.
If there is no difference of features with neighbors at small region near
position $\br$, they can be denoted by $\nabla\psi(\br)=0$ (or
$\nabla\phi(\br)=0$).
If there exists small tilting of phase angle at position $\br$ and an arbitrary
vector $\bA(\br)$ denote the difference between phase angles, the called
{\em covariant derivative} is given by $(\nabla-i\bA(\br))\psi(\br)=0$ (or
$\nabla\phi(\br)-A(\br)=0$).
If the covariant derivative vanishes (said to be flat or parallel by translated
in manifold theory) for all $\br$, the distribution of the field variables
$\psi(\br)$ would be a minimum solution of the integral
\begin{eqnarray} \label{eq:action}
  S=\int d\br\ |(\nabla-i\bA)\psi|^2,
\end{eqnarray}
for the connexion $\bA$.
A non-vanishing connexion $\bA$ occurs when there are strong competitive
behavior or inhibitory lateral interactions between neurons, and is related to
the emergence of a periodic pattern in cortical maps, such as the band patterns
in ocular dominance columns and the linear zones in orientation preference
columns, with the wavelength $\Lambda=2\pi/|\bA|$.
Fig.\ref{fig:macaque} shows the complete pattern of ocular dominance stripes
of a macaque monkey.
The orthogonality between the contour lines of feature map and the boundary
of cortical area is a property of minimal solutions in Eq.(\ref{eq:action}).
From the condition $\delta S/\delta\phi\sim 0$ or $\nabla^2\phi\sim 0$ for
$\psi=e^{2i\phi}$ with the preferred angle $\phi$, the normal component of
$\nabla\phi$ vanishes at the area boundary since the integral along a narrow
rectangular loop over the area boundary $\oint_C\nabla\phi\cdot d\hat{n}$
vanishes due to the divergence theorem.
Such perpendicularity with the area boundary is also manifested in other static
field solutions, such as the magnetic field.

\begin{figure}[t] 
\begin{minipage}[b]{5cm}
  \includegraphics[width=5cm]{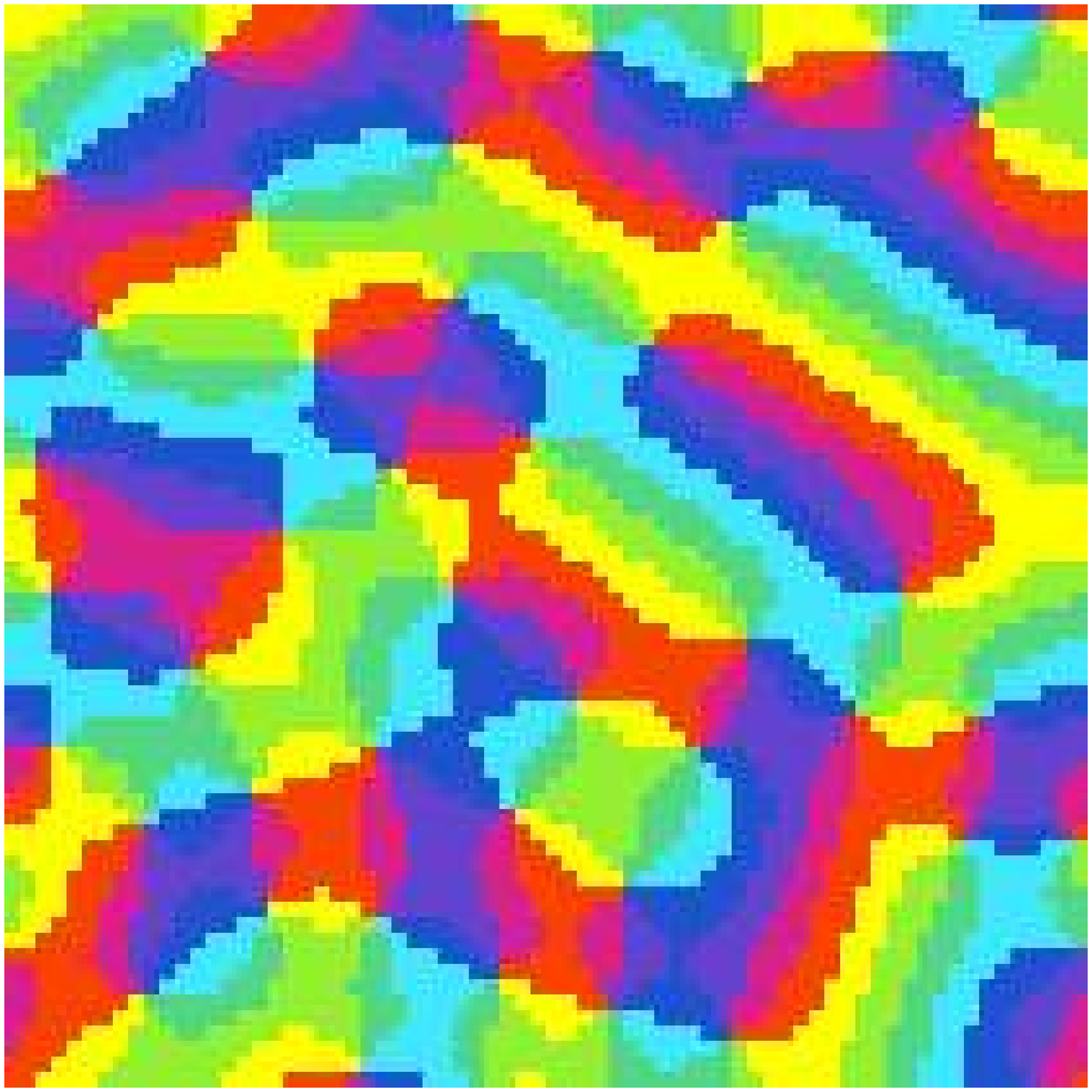}
\end{minipage}
\ \ 
\begin{minipage}[b]{0.65cm}
  \includegraphics[width=0.65cm]{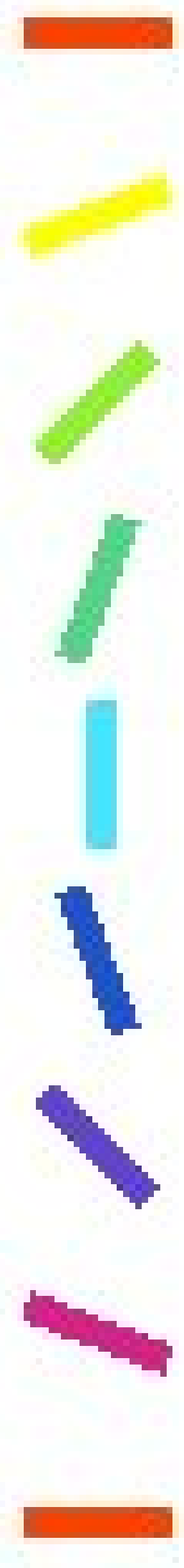}
\end{minipage}
\caption{ \label{fig:orientation_map}
  The simulation result of orientation map formation.
  The orientation maps have $U(1)$ (or $O(2)$) symmetry and the major
  characteristics of the developed map can be predicted using only the symmetry
  properties.
}
\end{figure}

The symmetry property also helps to predict the energy function of the cortical
map formation.
%The major features in cortical maps are universal and can be understood through
%the experience in other physical systems.
%Other orientation development models should also satisfy the energy form in spite
%of each different interaction rules employed.
For example, the features of orientation preference columns in the visual
cortex have $U(1)$ (or $O(2)$) symmetry.
However we perform a rotation in all the preferred angles through same angle
($\phi\rightarrow\phi+\chi$ - called `global' gauge transform), the energy of
orientation map formation should remain invariant.
The rotation angle $\chi$ can have a dependency on position $\br$, the called
`local' gauge transform, and the energy in a continuum approximation may take
the form in Eq.(\ref{eq:continuum}) or Eq.(\ref{eq:action}) with
$\bA=\nabla\chi(\br)$.

\section{Description of detailed neural interactions}
%If we consider the selective response to input signals due to the connections
%between neurons within a cluster, the attributes of a block of neurons can be
%represented following the basic neural network architectures, what Kohonen
%(1977) labeled {\em heteroassociation}.
%The cortical modification models at low level (or single-cell models) suggest
%more physical features of neural interactions and the biologic foundation of
%more abstract models.
The description of neural dynamic at a high level also should be based both on
neuroscience and informatics.
One important principle for neural plasticity is the Hebbian rule : two
simultaneously active neurons on either side of a connection increases the
weight of that connection is increased~\cite{Hebb1949}.
The simple Hebbian plasticity rule in a single neuron consists of inputs $\bu$
and weights $\bW$ takes the form
\begin{eqnarray} \label{eq:Hebbian}
  \Delta\bW(t)\propto y(t)\bu(t)
\end{eqnarray}
for the output $y=f(\bW\bu)$ with the activation function $f$ of simple cell.
In intracortical connected networks, the input becomes the summation of the
current from input and neighbor cells.
The output of neuron at $i$-th site becomes
\begin{eqnarray} \label{eq:recursive}
  y_i=f(v_i+\sum_j J_{ij}y_j)
\end{eqnarray}
for $v_i=W_{i\alpha}u_\alpha$ and the recurrent weight matrix $\bJ$.
In a energy model, synaptic plasticity rule is regarded as the negative
gradient of an energy (often objective, error or cost function) defined as a
function of $\bW$ :
\begin{eqnarray}
  \Delta\bW\propto-\frac{\partial E[\bW]}{\partial\bW}.
\end{eqnarray}
Because of the nonlinearity of the activation function and the recursive form
in Eq.(\ref{eq:recursive}), the energy used to be approximated depending on
models.
For example, with the assumption of $y_i=f(v_i+\sum_j J_{ij}v_j)$ and a series
expression of activation function $f(v)=\sum_\ell a_{\ell+1}v^\ell$, the energy
is obtained by
\begin{eqnarray} \label{eq:simple_energy}
  E[\bW]=-\sum_\ell\frac{a_\ell}{\ell}D_{i_1\cdots i_\ell}^{(\ell)}
    Q^{(\ell)}_{\alpha_1\cdots\alpha_\ell}
    W_{i_1\alpha_1}\cdots W_{i_\ell\alpha_\ell},
\end{eqnarray}
where
\begin{eqnarray}
  D^{(\ell)}_{i_1\cdots i_\ell}&=&(\delta_{i_1i_2}+J_{i_1i_2})\cdots
   (\delta_{i_{\ell-1}i_\ell}+J_{i_{\ell-1}i_\ell}) \nonumber \\
  &=&D^{(2)}_{i_1i_2}\cdots D^{(2)}_{i_{\ell-1}i_\ell}
\end{eqnarray}
is the functional tensor of rank $\ell$ and
\begin{eqnarray}
  Q^{(\ell)}_{\alpha_1\cdots\alpha_\ell}=
    \langle u_{\alpha_1}\cdots u_{\alpha_\ell}\rangle_\mD
\end{eqnarray}
is the input correlation tensor of rank $\ell$.
$\langle\ \cdot\ \rangle_\mD$ denotes the average over input data set $\mD$.
This energy based on the basic Hebbian rule is adjusted again depending on the
characteristic of synaptic plasticity rules~\cite{Fregnac1998}.
For example, the covariance plasticity rule replaces the input correlation
function $\bQ^{(\ell)}$ with rank $\ell$ as the input covariance function
\begin{eqnarray}
  C^{(\ell)}_{\alpha_1\cdots\alpha_\ell}=\langle
    (u_{\alpha_1}-\langle u_{\alpha_1}\rangle_\mD)\cdots
    (u_{\alpha_\ell}-\langle u_{\alpha_\ell}\rangle_\mD)\rangle_\mD.
\end{eqnarray}
In the FBM representation of simple cell model, feedforward synaptic weights
$\bW_{i\alpha}$ is replaced as field variables $\psi_\alpha(\br_i)$, then the
energy in Eq.(\ref{eq:simple_energy}) satisfies the form of energy in
Eq.(\ref{eq:expansion}).
For efficient description of dynamics, the energy is decomposed into the
functions of transformed field variables.
Because of the anisotropy in input correlation $\bQ$ (often in neighbor
activity $\bD$), the symmetry between components is broken and the effective
dynamics can be described with a few of dominant components.
The consequence of the anisotropy in neighbor activity between feature
components is explored in the case of the anisotropy between orientation and
ocular dominance columns~\cite{Cho2004B}.
%For the orientation preference columns, the prominent pattern are the oriented
%images with low frequency.

In a complex cell model, the features of neurons relate to the synapses within
a columnar module.
The columnar module is a kind of adaptive neural network systems and the
modulation of its functional attributes involves intricate changes in synaptic
weights.
An effective assumption is that the output of a columnar module is one of
the proper states of the functional and will change following afferent
signals.
For the currents from input and neighbor cells and a linear activation function,
the change in the proper state or the output of a columnar module is then
\begin{eqnarray}
  \Delta\by_i\propto\bv_i+\sum_jJ_{ij}\by_j
\end{eqnarray}
for the input $\bv_i$ to the columnar module at position $\br_i$ and the energy
averaged over inputs is obtained by
\begin{eqnarray}
  E[\by]=-\sum_i\langle\bv_i\rangle_\mD\by_i
    -\frac{1}{2}\sum_{i,j}J_{ij}\by_i\by_j.
\end{eqnarray}
In the FBM representation, the output with multivariable is replaced by field
variables :
\begin{eqnarray}
  E[\psi]&=&-\sum_iB_i\psi_i-\frac{1}{2}\sum_{i,j}J_{ij}\psi_i\psi_j
\end{eqnarray}
or
\begin{eqnarray}
 \lefteqn{E[\psi]=-\sum_i B(\br_i)\psi^\dagger(\br_i)} \\
  &&-\frac{1}{4}\sum_{i,j}J(\br_i,\br_j)\left\{\psi(\br_i)^\dagger\psi(\br_j)
    +\psi(\br_i)\psi(\br_j)^\dagger\right\}, \nonumber
\end{eqnarray}
where a functional vector $B(\br_i)=\langle \bv_i\rangle_\mD$ is the linear
average over inputs.
%The term for neighbor interactions in the FBM methods takes the exchange energy form
%Indeed the mathematical frameworks and formulas in the FBM methods resemble those
%in statistical quantum field theory. 
If we assume $\psi^\dagger$ and $\psi$ are creation and annihilation operators,
the term $\psi(\br_i)J(\br_i,\br_j)\psi^\dagger(\br_j)$ can be regarded as the
description of phenomena that a created activity at position $\br_j$ is
translated with kernel $J$ and annihilated at position $\br_i$.

A series of physiological experiments showed that the synaptic plasticity comes
from a redistribution of the available synaptic efficacy, not an increase in
the efficacy~\cite{Markram1996,Fregnac1998}.
In other words, the neural plasticity at the network level can be understood
as the pursuit of increment in the probability of reactivity with bounded total
synaptic strength for environmental experience.
With the expectation of a automatic normalization of synaptic weights,
%to a single neuron for simple cell model (or within a columnar module for complex cell model),
the norm of field variables $|\psi|$ used to be constrained to be constant.
In this sense, the neural dynamics with functional modularity may be described
by the slight shift in the internal phase per activity following afferent
signals.
Sometimes the normalization constraint is not imposed and involved in the
plasticity rule with subtractive normalization~\cite{Oja1982}.
For the energy function of the form
\begin{eqnarray}
  E[\psi]=a\psi^2-b\psi^4,
\end{eqnarray}
the stability of synaptic weight can be achieved due to the relaxation of
$|\psi|^2$ to its equilibrium value.

\begin{figure}[t] 
\begin{minipage}[b]{4cm}
  \includegraphics[width=4cm]{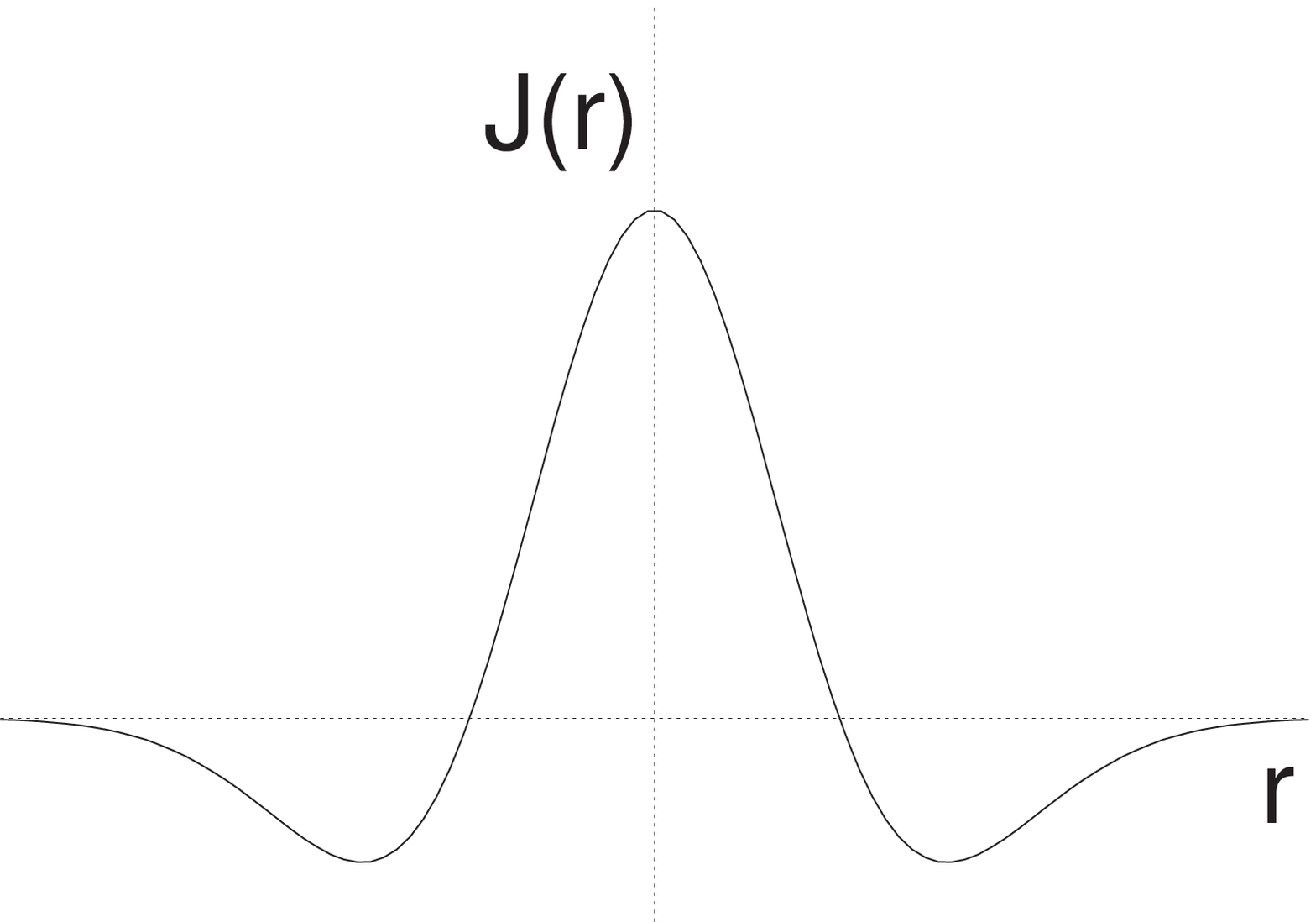} \\
 (a) Lateral feedback control of activity
%\\ \ \\
\end{minipage}
\ 
\begin{minipage}[b]{4cm}
  \includegraphics[width=4cm]{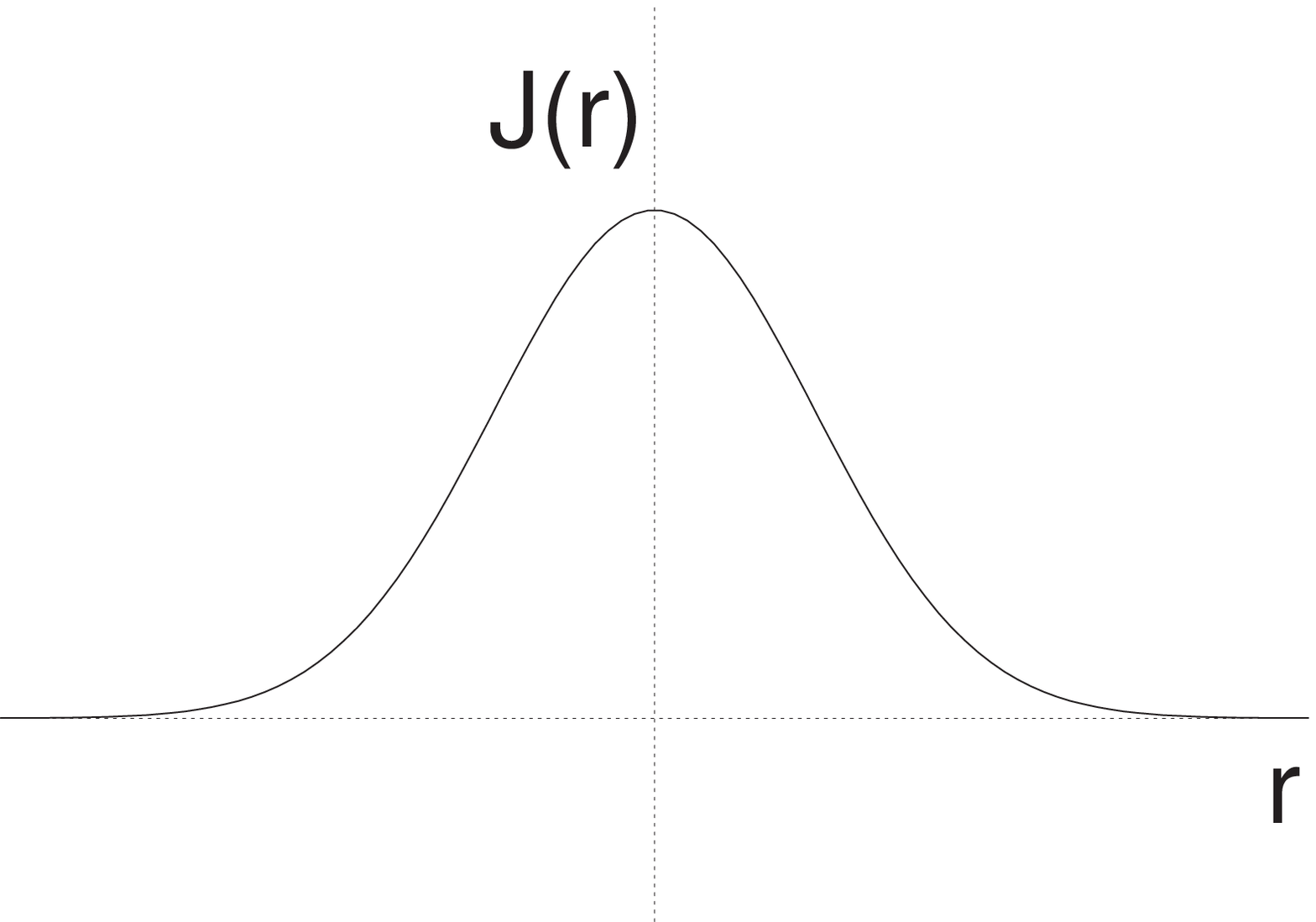} \\
  (b) Lateral control of plasticity
\end{minipage}
\caption{ \label{fig:control}
  The two types of neighbor interaction functions and control mechanisms.
  (a) The lateral interaction models adopt lateral activity control and the
  activation kernel, usually so-called ``Mexican hat'' function (positive
  feedback for close distance and negative for longer distance).
  (b)  The plasticity control with nonnegative kernel requires feedforward
  competition (or feedforward normalization of activity over networks).
  The elastic-net model assumes the nearest neighbor interactions (or elastic
  force), whereas the SOFM algorithm takes the neighbor function on Gaussian
  form with the hard competition (or winner-take-all activity).
}
\end{figure}

Another important mechanism expected in neural computation is the enhancement
of neural activity depending on correspondence to input.
A possible enhancement modulation is the restriction on the sum over the
activity by subtractive normalization.
With a simple nonlinear form $x+\eta x^2$, the external source term with
enhanced afferent signals becomes that
%depending on the conformity is that
\begin{eqnarray} \label{eq:enhancement}
  \langle\bv_i'\rangle_\mD\psi_i&=&\left\langle \bv\frac{\rho_i(1+\eta\bv\psi_i)}
    {(1/\rho)\sum_j\rho_j(1+\eta\bv\psi_j)}\right\rangle_\mD\psi_i \nonumber \\
  &\simeq&\langle\bv_i\rangle_\mD\psi_i+\frac{1}{2}\sum_j S_{ij}\psi_i\psi_j
\end{eqnarray}
for $\bv_i=\rho_i\bv$ and $\rho=\sum_i\rho_i$ with the stimuli strength
$\rho_i$ at position $\br_i$.
The scattering function with a input data set $\mD$ is defined as
\begin{eqnarray} \label{eq:scattering}
  S_{ij}=2\eta\langle v_iv_j\rangle_\mD(\delta_{ij}-1)
\end{eqnarray}
for the enhancement (or competition) parameter $\eta$.
In the FBM representation, the scattering function describes the feedforward
competition process in the competitive Hamiltonian models, such as the
elastic-net model and the SOFM algorithm.
%For hard competition with large $\eta$, the network accomplishes the
%``winner-take-all'' process.
%In fact, a priori enhancement of afferent signals is achieved when the
%conformity between neural feature and input signal is determined by the
%connectivity with the incentive cells (or extrinsic coding type).
In fact, for an intrinsic coding type, network cannot tell which neurons match
mostly with input signal a priori and the winner has to be determined after
lateral inhibitory activity.
The competitive Hebbian models require a normalization control of response or a
priori decision of winner and depict the feature vectors in visual cortex
through the connectivity between visual cortex and retinas (or
LGNs)~\cite{Scherf1999}.
%Another important role of synaptic normalization in cortical map development is the competition.
% in afferent signals is equivalent to those of the lateral inhibitory activity.
The lateral activity function $J(\br_i,\br_j)$, the connectivity between
neurons (or columnar modules) at position $\br_i$ and $\br_j$ within a cortex
area, has two types according to the control mechanisms~\cite{Kohonen1995}.
In the case of the lateral feedback control (which Kohonen called the
activity-to-activity kernel), the lateral activity function $\bJ$ is regarded
to be excitatory for short distance and inhibitory for long distance with the
so-called Mexican hat type (Fig.\ref{fig:control}a).
Whereas in the case of the lateral control of plasticity (or the
activity-to-plasticity kernel), the lateral interaction is nonnegative and may
take the Gaussian form (Fig.\ref{fig:control}b).
The competitive Hebbian models take the lateral control of plasticity, that
means there is no negative value in $\bJ$, and the scattering function $\bS$
from afferent signal enhancement has an effect of inhibitory activity.

%We can consider also the interactions with higher powers and take the general
%energy form as Eq.(\ref{eq:expansion}).
%Note that the actual forms of interaction functions depend on interaction
%mechanisms and the quadratic interaction term $D(\bx,\by)$ need not always be
%the neighborhood function $J(\bx,\by)$.

Now we employ the concepts of thermodynamic into neural dynamics.
In some classes of neural network models, such as Boltzmann machine, the 
input-output is assumed to be stochastic.
Once a stochastic neural network has converged to an equilibrium state, the
probability distribution characterizing $\psi$ is expected to obey the
Boltzmann distribution
\begin{eqnarray}
  P[\psi]=\frac{\exp(-E[\psi])}{Z}
\end{eqnarray}
for the partition function
\begin{eqnarray}
  Z=\sum_\psi\exp(-E[\psi]).
\end{eqnarray}
In neural processing architecture, the notion of entropy or free energy is put
into practice ahead for the purpose of informatics.
Compared to deterministic firing models, an expected advantage in stochastic
neural network models is to escape from poor locally optimal configurations
through probabilistic evolution.
Moreover, there are several reasons that the stochastic behavior should be
indispensable process in neural networks.
In view of learning rules, it is natural that neural states are occupied
with features corresponding to frequent inputs (the {\em coarse coding}
principle).
On the other hand, it is efficient for a neural network to avoid the occupation
with a few of features, so that an object is coded by a small population that
is active for an event (the {\em sparse coding} principle).
%Besides the competitive or inhibitory activity, thermodynamic behavior in neural
%networks tends to achieve the sparseness.
It is usual that the cost function in unsupervised learning algorithm is
similar to the Helmholtz free energy that
\begin{eqnarray} \label{eq:Helmholtz}
  F=E-TS,
\end{eqnarray}
where the parameter $T$ is considered just as a positive constant that
determines the importance of the second term relative to the first.
%In a Hebbian development model, the energy term $E$ functions neurons to
%possess features corresponding frequent inputs in addition to the neighbor
%ordering, whereas the entropy term compels neurons to avoid occupying a
%common feature state.
In learning rules, the energy term is illustrated by a measurement how well the
code describes the input  data or carry the informations :
\begin{eqnarray} \label{eq:E}
  %E&=&(1/N)\sum_{i}\langle P(\psi_i|\bv)\rangle_\mD \nonumber \\
  E&=&(1/N)\sum_{i}\sum_{\bv\in\mD}
    P(\psi_i|\bv)P(\bv|\mD) \nonumber \\
  &=&(1/N)\sum_{i}P(\psi_i|\mD) \\
  &=&\sum_{\psi}P(\psi)P(\psi|\mD), \nonumber
\end{eqnarray}
%In a network with the input-output stochastic relationship,
where  a distribution $P(\psi|\mD)$ is the average over the probability that
input $\bv\in\mD$ generates output $\psi$.
In Hebbian development models, this energy term can be considered as an
external source term, that is the average over the product between feature
state and external signals, $-B\psi$ in a complex cell model (or
$-\psi\bQ^{(2)}\psi$ in a simple cell model) as well.
%However, if the neural dynamics is described by $E=V(\psi)$, the solutions
%indicate the collapse of whole neurons to single feature state with the maximal
%probable experience.
%We can expect that the observed cortex maps {\em in vivo} are aparted from the
%equilibrium state because the relaxation process in neuron systems is very
%slow, and they will reach to single state finally.
%The {\em minimum description length} (MDL) principle~\cite{Rissanen1989}, for
%example, finds a method of coding each input data that minimizes the total cost
%of communicating the input data to a receiver.
%The energy is described as $P(\psi)$ is the probability of the feature state
%$\psi$ in the cortex area or the prior probability of the model $\psi$.
In learning rules, the entropy term assesses the sparseness of the code by
assigning a cost depending on how the activity is distributed.
According to Shannon's coding theorem, the amount of information is defined by
\begin{eqnarray} \label{eq:S}
  S=-K\sum_{\psi} P(\psi)\ln P(\psi)
\end{eqnarray}
where $K$ is a positive constant and $-\ln P(\psi)$ is the cost of the code,
the number of bits required to communicate the code.
%The expression for $E$ and $S$ in Eq.(\ref{eq:E}) and Eq.(\ref{eq:S}) remind
%the Helmholtz's free energy in density matrix formulation,
%\begin{eqnarray} F[\rho]=\mbox{Tr}\ \rho\left\{H+k_BT\ln\rho\right\} \end{eqnarray}
%for the density matrix $\rho$ with $K$ being identified as Boltzmann's constant $k_B$.
The connections between information theory and statistical mechanics are
rigorously investigated~\cite{Jaynes1957A,Jaynes1957B,Grandy1997}.
%From the point of the learning algorithm, the probabilistic decision neural
%networks such as {\em Boltzmann machines}~\cite{Hinton1983} have been suggested.
However, there is some hardship to apply the statistical mechanism to the
phenomena in the real brain.
Since neural process, comprehends the dynamics at various spatial and temporal
levels, is essentially dynamical and non equilibrium phenomenon.
For example, the relaxation process in cortical map formation is very slow and
the observed maps often do not satisfy the equilibrium criteria.
The map formation in visual cortex occurs concentrately for several weeks or
months after birth, during a so-called critical period.
In observed orientation preference maps, the non-uniforming directions of
gradient ($\nabla\phi_\parallel\neq\mbox{const}$, however
$|\nabla\phi_\parallel|\simeq\mbox{const}$ for the longitudinal component
$\phi_\parallel$) and non-vanishing singular points (or pinwheels) indicate
that the system may be frozen during the relaxation process~\cite{Cho2004A}.
% the perpendicularity with
%area boundary ($\nabla^2\phi_\parallel\sim 0$ for the longitudinal component
%$\phi_\parallel$) is achieved, but except
%The relaxation process in cortical dynamics is very slow and the observed maps
%in adult are expected to be stopped in process.
%We think the neural networks should be treated as aparted from but in
%relaxation to the equilibrium state.
%we can guess how far they apart from the equilibrium state.

\section{Application to visual map formation models} \label{sec:application}
According to the studies of the statistical structure of natural images, the
response properties of visual neurons, the spatially localized and oriented,
are considered to be due to the efficient coding of natural
images~\cite{Olshausen1996}.
Oriented bar or grid patterns are the most probable activity and the feature
with $O(2)$ (or $U(1)$) symmetry components is a meaningful representation in
orientation columns.
With ocular dominance columns, the total feature can be expanded to $O(3)$
symmetry components with the restriction of synaptic normalization within
each column.
Therefore, the conventional spin vector $(S^x,S^y,S^z)$ can serve as a useful
representation of the feature states with the preferred orientation
$\phi=(1/2)tan^{-1}(S_x/S_y)$.
%Among the phenomena at the cortical level, the primary visual map formation is
%one of the most investigated problems with various proposed models, most of
%which are described in the high- or low-dimensional feature vector representation.
The proposed models of visual map formation are based on so various mechanisms.
%which factor causes the bifurcation to a inhomogeneous state.
We rewrite four widely used visual map development models in term of FBM
representation classified by their effective interaction terms that
\begin{itemize}
\item[(A)] Lateral interaction models : $\bD = \bJ$
\item[(B)] Recursive interaction models : $\bD = (\bI-\bJ)^{-1}$
\item[(C)] Elastic-net model : $\bD = \bJ + \bS$
\item[(D)] SOFM algorithm : $\bD = \bS\bJ$.
\end{itemize}
The lateral activity function $J(\br_i,\br_j)$ in the elastic-net model and
the SOFM algorithm is taken to be all nonnegative, the two-point interaction
function $D(\br_i,\br_j)\simeq D(|\br_i-\br_j|)$ takes the Mexican hat type for
all cases owing to the scattering funciton $S(\br_i,\br_j)$.
%The competitive Hebbian models require the feedforward competition and the
%nonnegative neighborhood function (plasticity control kernel) with the
%enhancement of activity in Eq.(\ref{eq:enhancement}).
Periodic patterns, such as linear zones in orientation preference columns or
parallel bands in ocular dominance columns, can develop when there are abundant
negative values in $D(\br_i,\br_j)$ so that $\tilde{D}(q)$ in Fourier space has
a non-vanishing minimum point $q^\ast$ with the wavelength
$\Lambda=2\pi/q^\ast$.
%We show that the elastic-net model and the lateral interaction models can be
%described by common energy form in Eq.(\ref{eq:linear}) with
%$D(r)=h_+(r)-h_-(r)$ for positive functions $h_+$ and $h_-$.
%This result means that two models have equivalent effective interactions and
%share statistical properties, in spite of their different control mechanisms.
%The emergence of columnar patterns are possible when $h_-$ is relatively larger
%than $h_+$.
%The lateral interaction models consider $h_-$ as the lateral inhibitory
%interaction term whereas the elastic-net model suggest it as the correlated
%external stimuli term.

\subsection{Lateral Interaction Models}
A simple cell model of the visual map development uses the high-dimensional
feature vector coding for the strength of the connection from each cortical
location to each retinal (or LGN) location.
The synaptic plasticity depends on the average over the activities of competing
inputs, which are left and right eyes for ocular dominance columns or ON-center
and OFF-center cells for orientation preference columns.
For a linear activation function or $f(v)=v$, the energy in
Eq.\ref{eq:simple_energy} becomes that
\begin{eqnarray}
  E[\bW]=-\frac{1}{2}\sum_{i,j}\sum_{\alpha,\beta}
    (\delta_{ij}+J_{ij})Q^{(2)}_{\alpha\beta}\bW_{i\alpha}\bW_{j\beta}.
\end{eqnarray}
This energy is decomposed with the (globally) transformed synaptic weights into
the sum and difference :
%The synaptic weights are represented by their transformation into the sum and difference :
\begin{eqnarray}
  \bW_+=\bW_R+\bW_L & \mbox{and} & \bW_-=\bW_R-\bW_L
\end{eqnarray}
for ocular dominance columns or similarly $\bW_\pm=\bW_{ON}\pm\bW_{OFF}$ for
orientation columns.
In a pixel-based representation for orientation columns, oriented patterns with
low-frequency compose a dominant feature space and the energy is decomposed
with the locally transformed weights as well.
Therefore, the energy as the function of field variables $\psi$ in a
transformed and reduced feature space is that
\begin{eqnarray}
  E[\psi]=-\frac{1}{2}\sum_{i,j}D(\br_i,\br_j)\psi(\br_i)\psi(\br_j)
\end{eqnarray}
for $\bD=\bI+\bJ$.
The input correlation matrix $\bQ^{(2)}$ is ignored, because the frequency of
the dominent features in inputs is regarded to be the same or the two-point
activity function $\bD$ comprises this.
The term $-\frac{1}{2}\sum_i\psi(\br_i)^2$, related to the self-relaxation
term, does not effect influence on the typical spacing of an emergent columnar
pattern.
In a complex cell model, the simplest is given by the summation of the neighbor
interactions and the external stimuli terms as
\begin{eqnarray} \label{eq:linear}
  E[\psi]=-\frac{1}{2}\sum_{i,j}D(\br_i,\br_j)\psi(\br_i)\psi(\br_j)-\sum_iB(\br_i)\psi(\br_i)
  %E[\psi]=-\frac{1}{2}\sum_{i,j}D_{ij}\psi_i\psi_j-\sum_iB_i\psi_i
\end{eqnarray}
for $\bD=\bJ$.
The external stimuli $B(\br_i)$ is considered to be constant or vanishing.
Therefore, the form of the lateral activity function $\bJ$ determines the
typical appearance of developed feature map for both cases.
In lateral interaction models, $\bJ$ is taken as the {\em activation kernel},
or Mexican hat function (positive feedback in the center, negative in the
surroundings).
For example, a well-known Mexican hat function, the called difference of
Gaussians (DOG) filter, is described as
\begin{eqnarray}
  J(\br_i,\br_j)=\varepsilon\left(e^{-|\br_i-\br_j|^2/2\sigma_1^2}
    -ke^{-|\br_i-\br_j|^2/2\sigma_2^2}\right)
\end{eqnarray}
where $k$ is the strength of inhibitory activity.
Another example of Mexican hat function modified from a wavelet is given by
\begin{eqnarray} \label{eq:wavelet}
  J(\br_i,\br_j)=\varepsilon\left(1-k\frac{|\br_i-\br_j|^2}{\sigma_l^2}\right)
    e^{-|\br_i-\br_j|^2/2\sigma_l^2}
\end{eqnarray}
for the lateral cooperation range $\sigma_l$.
If the strength of inhibitory activity $k$ is larger than threshold $k_c$
($=1/4)$, $\tilde{D}(q)$ has a non-vanishing maximum point at
$q^\ast=(1/\sigma)\sqrt{4-1/k}$~\cite{Cho2004A}.

\subsection{Recursive interaction models}
For a linear activation function, the output in Eq.(\ref{eq:recursive}) becomes
\begin{eqnarray}
  y_i=v_i+\sum_jJ_{ij}v_j+\sum_{j,k}J_{ij}J_{jk}v_k+\cdots,
\end{eqnarray}
which is the summation of recursive recurrents.
The energy as the function of synaptic weights is obtained that
\begin{eqnarray} \label{eq:correlation-based}
  E[\bW]=-\frac{1}{2}\sum_{i,j}\sum_{\alpha,\beta}D_{ij}Q^{(2)}_{\alpha\beta}
    \bW_{i\alpha}\bW_{j\beta},
\end{eqnarray}
where the two-point interaction function is
\begin{eqnarray}
  \bD=\bI+\bJ+\bJ^2+\cdots=(\bI-\bJ)^{-1}
\end{eqnarray}
and the real parts of the eigenvalues of $\bJ$ are expected to be less than $1$.
Eq.(\ref{eq:correlation-based}) is a simple modified equation of Miller's
correlation-based learning models~\cite{Dayan2001}.
In the original representation by Miller {\em et al.}, the input stimuli term
is described by an arbor function, expressing the location and the overall size
of the receptive fields~\cite{Miller1989,Miller1994}.
The two-point interaction function $\bD$ takes the Mexican hat type and the
wavelength of visual pattern is determined by the peak of $\tilde{D}(q)$ in
the analysis by Miller as well~\cite{Miller1998}.

%The ocular dominance development model by Miller {\em et al.} uses a
%high-dimensional feature vector coding for the strength of the connection from
%each cortical location to each retinal (or LGN) location.
%The correlation-based models is based on the synaptic plasticity depending on
%the correlations among the activities of competing inputs, which are left and
%right eyes for ocular dominance columns or ON-center and OFF-center cells for
%orientation preference columns

\subsection{Elastic-Net Model}
The elastic-net model is described by an iterative procedure
with the update rule :
\begin{eqnarray} \label{eq:elastic}
  \Delta\Phi(\br_i)&=&\alpha\sum_{|\br_i-\br_j|=a}(\Phi(\br_i)-\Phi(\br_j))
    \nonumber \\
  &+&\beta(\bV-\Phi(\br_i))\frac{e^{-|\bV-\Phi(\br_i)|^2/2\sigma_s^2}}
    {\sum_j e^{-|\bV-\Phi(\br_j)|^2/2\sigma_s^2}}, \ \ \ \ 
\end{eqnarray}
where a feature vector in the low-dimensional representation is
\begin{eqnarray}
  \Phi(\br)&=&(r_x,r_y,q\sin(2\phi(\br)),q\cos(2\phi(\br)),z(\br)) \nonumber \\
  &=&(\br,\psi(\br)) \nonumber
\end{eqnarray}
for the retinal location $\br=(r_x, r_y)$, the preferred orientation
$\phi(\br)$, the degree of preference for that orientation and the ocular
dominance $z$~\cite{Durbin1990,Erwin1995}.
At each iteration, a stimulus vector $\bV=(\br_v,\bv)$ is chosen at random
according to a given probability distribution.
The first term in Eq.(\ref{eq:elastic}) denotes the elastic force or the
excitatory interactions between the nearest-neighbors, and the second term
implies the normalized stimuli distributed around an activity center.
Functional Taylor expansion of the right hand side after dropping all nonlinear
terms leads to
\begin{eqnarray}
\lefteqn{\Delta\psi(\br_i)=\alpha \sum_{|\br_i-\br_j|=a}
  \big\{\psi(\br_i)-\psi(\br_j)\big\}} \\
&&-\beta\psi(\br_i)+\frac{\beta a^4}{4\pi^2\sigma_s^6}
  \sum_j\langle\bv_i\bv_j\rangle_\mD\big\{\psi(\br_i)-\psi(\br_j)\big\} \nonumber
\end{eqnarray}
where the stimulus at position $\br_i$,
\begin{eqnarray} \label{eq:scatter}
  \bv_i=\bv e^{-|\br_i-\br_v|^2/2\sigma_s^2}
\end{eqnarray}
is distributed in a gaussian form with the activity center $\br_v$ and the
feedforward cooperation range $\sigma_s$.
The correlation between the external stimuli at position $\br_i$ and $\br_j$ is
obtained by
\begin{eqnarray} \label{eq:correlation}
  \langle\bv_i\bv_j\rangle_\mD&=&\langle v^2\rangle_\mD
    \sum_{\br_v}\ e^{-|\br_i-\br_v|^2/2\sigma_s^2}
    \ e^{-|\br_j-\br_v|^2/2\sigma_s^2} \nonumber \\
  &\simeq&(\pi\sigma_s^2/a^2)\langle v^2\rangle_\mD
    e^{-|\br_i-\br_j|^2/4\sigma_s^2}.
\end{eqnarray}
%as follows the results of Hoffs\"{u}mmer {\em et al.}~\cite{Hoffsummer1995}.
Therefore, the effective energy of the elastic-net model can be represented
through the form in Eq.(\ref{eq:linear}), where the two-point interaction
function is given by
\begin{eqnarray}
  \bD=-\beta\bI+\bJ+\bS.
\end{eqnarray}
The lateral activity function becomes
$J(\br_i,\br_j)=\alpha\delta(|\br_i-\br_j|-a)$ or the Laplacian operator in a
continuum limit.
The scattering function coincides with the form in Eq.(\ref{eq:scattering})
for $\eta=\beta a^4/8\pi\sigma_s^6$ or is obtained by
\begin{eqnarray} 
  %S(\br_i,\br_j)&=&\frac{2\eta}{N}(\delta_{ij}-1)\langle v_iv_j\rangle_\mD  \\
  %  &=&\beta\frac{\langle v^2\rangle_\mD}{2\pi\sigma_s^2}
  %  \left(2\pi\delta_{ij}-e^{-|\br_i-\br_j|^2/4\sigma_s^2}\right) \nonumber
  S(\br_i,\br_j)\simeq\frac{\beta}{\sigma_s^2}\langle v^2\rangle_\mD
    \left(\delta_{ij}-\frac{a^2}{4\pi\sigma_s^2}e^{-|\br_i-\br_j|^2/4\sigma_s^2}\right).
\end{eqnarray}
This result means that the scattering function $S(\br_i,\br_j)$ can act as an
kernel with inhibitory activity however the lateral activity function
$J(\br_i,\br_j)$ is nonnegative.
There are also interaction terms of higher power but the two-point interaction
function $D(\br_i,\br_j)$ determines the major characteristics of developed
feature maps.
We transform it to Fourier space and obtain
\begin{eqnarray}
  \tilde{D}(\bq)&=&-\beta+\tilde{J}(\bq)+\tilde{S}(\bq) \\
  &\simeq&-\beta-\alpha q^2+\frac{\beta}{\sigma_s^2}
    \langle v^2\rangle_\mD\left(1-e^{-q^2\sigma_s^2}\right). \nonumber
\end{eqnarray}
It has a maximum at
\begin{eqnarray}
  q^\ast=\frac{1}{\sigma_s}\sqrt{\ln
    \left(\frac{\beta}{\alpha}\langle v^2\rangle_\mD\right)},
\end{eqnarray}
which corresponds to the analytic results from different
approaches~\cite{Hoffsummer1995,Scherf1999}.
%The maximum is positive for any $\sigma_s<\sigma_s^\ast$ where
%\begin{eqnarray}
%  \sigma_s^\ast=\sqrt{\langle v^2\rangle_\mD
%  -\alpha-\alpha\ln\left(\frac{\beta}{\alpha}\langle v^2\rangle_\mD\right)}.
%\end{eqnarray}
%The sequence bifurcation model.

\subsection{Self-Organizing Feature Map Algorithm}
In Eq.(\ref{eq:linear}), the interaction term $\psi J\psi$ denotes the exchange
of spontaneous spikes, created without external activity.
Spontaneous firings can occur in coupled nonlinear oscillators with small
dynamic fluctuations, which have been observed in some experiments~\cite{
Llinas2003,Creutzfeldt1995,Steriade1993,Tsodyks1999,Sanchez2000,Wilson1981}.
However, several experiments suggested that the organization of feature maps is
possible after the exposure to the external activity.
In this case, the probability of spontaneous firing are small ($J\ll v$), so
that the most intracellular interactions would be achieved by indirect currents
of external activities.
With the provoked interactions by external activities, we can take the
effective energy as
\begin{eqnarray} \label{eq:H_SOM}
  E[\psi]=-\left(\sum B\psi+\frac{1}{2}\sum\psi S\psi\right) 
    \left(\frac{1}{2}\sum \psi J\psi\right).
\end{eqnarray}
If $B(\br)\psi(\br)$ is constant for all position $\br$, the first term with
$\psi J\psi$ supports the lateral interaction models again.
In the Kohonen's SOFM algorithm, the lateral currents induced by feedforward
normalized stimuli are focused and the effective interaction term is given by
%ignores this term or assumes that $B(\br_i)=\langle v(\br_i)\rangle_\mD$ vanishes, it focus on
\begin{eqnarray}
  D(\br_i,\br_j)&=&\frac{1}{2}\sum_\br S(\br_i,\br)J(\br,\br_j).
\end{eqnarray}
Moreover, the SOFM algorithm requires the hard competition, the called
``winner take all'' (WTA) case.
As $\sigma_s$ approaches zero (or large $\eta$), the activity is localized only
around the winning neuron and the scattering function in Fourier
space becomes $\tilde{S}(\bq)\simeq\beta \langle v^2\rangle_\mD q^2$, the
Laplacian operator.
The lateral activity function in the SOFM approaches takes on the Gaussian form
$J(\br_i,\br_j)=e^{-|\br_i-\br_j|^2/2\sigma_l^2}$ for the lateral cooperation
range $\sigma_l$ (lateral plasticity control).
Therefore we obtain the two-point interaction function
\begin{eqnarray} \label{eq:FBM_Dq}
  \tilde{D}(\bq)=\frac{1}{2}\tilde{S}(\bq)\tilde{J}(\bq)
    =\pi \sigma_l^2\beta\langle v^2\rangle_\mD\ q^2e^{-q^2\sigma_l^2/2}
\end{eqnarray}
in Fourier space or
\begin{eqnarray}
  D(\br_i,\br_j)=\beta\langle v^2\rangle_\mD \left(1-\frac{|\br_i-\br_j|^2}
    {2\sigma_l^2}\right)e^{-|\br_i-\br_j|^2/2\sigma_l^2}
\end{eqnarray}
in real space.
This is the Mexican hat function in Eq.(\ref{eq:wavelet}) with $k=0.5$.
Eq.(\ref{eq:FBM_Dq}) has a minimum at
\begin{eqnarray}
  q^\ast=\sqrt{2}/\sigma_l,
\end{eqnarray}
which agrees with previous analytic
results~\cite{Wolf2000,Scherf1999,Obermayer1992} and always positive if
$\sigma_l>0$.
The Kohonen's SOFM algorithm reads to robust learning rules because it always
succeeds in achieving an array of different feature detectors or a columnar
pattern.

\section{Discussion}
The physical models of neural network based on neuroscience attempt to
interpret both physiologic phenomena and computational architectures.
In order to study the functional of the real brain, we need more adaptable
theories than the basic neural architecture with connectionism.
In this paper, we show that the neural process at the cortical level can be
described by using the conventional expressions in statistical physics.
As we showed in visual map formations~\cite{Cho2004A,Cho2004B}, the collective
neural dynamics can be much alike well-known phenomena in the physical systems.
%More extended computational architectures are also possible in the neural
%models with functional modularity because of the higher dimensional attributes
%of the processing elements in networks.
%The neural dynamic models at higher levels also have to base on neuroscience
%and target to interpret both the physiologic phenomena and the computational
%architectures.
%The representation of neural dynamics at the cortical level is suited to understand
%the statistical and collective phenomena of neurons - multi-functional map
%organization or map differentiations, and cooperative computation, etc.
%More effective description of neural dynamics and computations with functional
%and columnar modularity have been suggested because the connections and
%interactions between neurons are very huge and complex in real brain.

In the assumption of neural network composed of columnar modules, we classify
the synaptic connection types and anticipate different functional characters in
computational processing.
(1) In the connectivity between close neurons within a columnar module, the
functional attributes of neurons and the associative memory is realized.
(2) By the connectivity between columnar modules within a cortex area, noted by
the lateral activity function or recurrent weight matrix $\bJ$, the networks
control laterally the output activity between neighbors.
(3) Via the connectivity between far apart neurons cross cortex areas, neurons
get driven-activity from external environment or other functional cortex areas.
The columnar modules become elements (or nodes) again with high dimensional
attributes in networks of neural networks.
If the recurrent weights matrix $\bJ$ is specified depending on the positions,
the connectivity between columnar modules also work in information coding.
The connectivity between columnar modules within or beyond cortex areas would
be strengthened also if there are much communications between them according to
the Hebbian rule, and there are some models holding the updating rule in the
recurrent weights matrix $\bJ$, such as the Goodall rule.~\cite{Goodall1960}.
We regard that the enhancement of connectivity between columnar modules proceed
to the efficient communications between neurons rather than information coding.
The consideration of minicolumn as a columnar module and the processing element
in network is optional.
The formation of structure in minicolumn is also due to the functional grouping
between neurons with similar interests, and expected to be certified with more
fundamental process at the cellular or molecular level.

%Like the linear analysis in other models, the direct interaction term is
%important in the determination of dominance feature of self-organizing map.
%Higher power interaction terms are possible if considering (1) the normalization
%of synaptic strength over networks, (2) indirect interactions between neurons,
%and (3) thermodynamic perturbative coupling by $g\psi^4$ term.
%The statistical property of the cortical map is revealed in the general energy
%formula at continuum limit such as Eq.(\ref{eq:continuum}).
%Using the Landau theory, the prediction of phase transition in neural systems
%would be possible also.

%Indeed, the interactions in neurons resemble those in the physical particles.
%Neurons (or electrons) receive spikes (or photons) from neighbor neurons (or
%electrons) and send them to neighbors again.
%After collision with spikes (or photons), the preferred state of neurons
%({\em or} the momentum or intrinsic phase of electrons) moves slightly to the
%driven stimuli.
%sensation, perception and memorization

Extraction of the significant features in the input data is the purpose of an
unsupervised learning rule and also expected to be a principle character of
artificial and physiologic neural networks.
The FBM representation method suggests how neurons find features from afferent
signals and build knowledgement at the cortical level.
An abstract representation of features in the FBM representation and a symmetry
breaking between feature components in progress is related to the learning
process in the neural network.
For example, difference looks of an object form a submanifold in pattern space
and the patterns of the object can be abstracted and decomposed in the
transformed and reduced feature space.
% such as angle or distance from viewpoint.

In view of dynamics, the essential factors in neural process are (1)
statistical structure of inputs, (2) attractive or repulsive interactions
between neighbor neurons, and (3) stochastic behavior of neurons.
In this paper, we did not fully apply thermodynamic mechanics into neural
process.
There are some models which contain thermodynamic approach.
The basic ingredients of Tanaka's Potts spin models are those of the lateral
interaction models but he took a probabilistic evolution rather than a energy
gradient flow~\cite{Tanaka1989,Tanaka1990A,Tanaka1990B,Tanaka1991A,Tanaka1991B}.
Piepenbrock presented a model which uses the effect of stochastic behavior in
neural network as a competition process~\cite{Rao2002}.
However there is no lateral inhibitory activity or feedforward competition,
thermodynamic effect can make a network to have a columnar structure with a
thermal excitation at low temperature.
We expect the stochastic behavior of neurons can be the connection between the
physical neural dynamic models and the neural network models originated from
learning theory and an essential factor in comprehension of systematic
ordering-disordering or bifurcation problems in the real brain.
Moreover, we expect that the theoretic experience in physics can offer more
intuitive appreciation of the physiologic phenomena at higher level and
sophisticated mechanisms in computational architecture.

This work was supported by the Ministry of Science and Technology and the
Ministry of Education.

\bibliography{fbm}

\end{document}